\documentclass{aa}
\usepackage{graphicx}
\usepackage{txfonts}
\usepackage{natbib}
\usepackage{color}
\usepackage{soul}
\usepackage[colorlinks, allcolors=blue]{hyperref}

\begin{document}

\title{IRIS burst properties in active regions\thanks{Associated data available at: 10.5281/zenodo.7016311}}

\author{C. J. Nelson$^{1}$, L. Kleint$^{2,3}$}

\offprints{chris.nelson@esa.int}
\institute{$^1$European Space Agency (ESA), European Space Research and Technology Centre (ESTEC), Keplerlaan 1, 2201 AZ, Noordwijk, The Netherlands.\\
$^2$University of Geneva, 7, route de Drize, 1227 Carouge, Switzerland.\\
$^3$Astronomical Institute of the University of Bern, Sidlerstrasse 5, 3012 Bern, Switzerland}

\date{}

\abstract
{Interface Region Imaging Spectrograph (IRIS) bursts are localised features thought to be driven by magnetic reconnection. Although these events are well-studied, it remains unknown whether their properties vary as their host active regions (ARs) evolve.}
{In this article, we aim to understand whether the measurable properties (e.g. frequency, area, location, spectral characteristics) of IRIS bursts are consistent during the evolution of their host ARs.}
{We study 42 dense 400-step rasters sampled by IRIS. These rasters each covered one of seven ARs, with each AR being sampled at least four times over a minimum of 48 hours. An automated IRIS burst detection algorithm is used to identify IRIS burst profiles in this work. Data from the Solar Dynamics Observatory's Helioseismic and Magnetic Imager are also used to provide context about the co-spatial line-of-sight magnetic field.}
{Of the rasters studied, 36 ($86$ \%) were found to contain IRIS burst profiles. Five ARs (11850, 11909, 11916, 12104, and 12139) contained IRIS burst profiles in each raster that sampled them whilst one AR (11871) was found to contain no IRIS burst profiles at any time. A total of 4019 IRIS burst profiles belonging to 752 connected objects, which we define as parent IRIS bursts, were studied. IRIS burst profiles were only detected within compact regions in each raster, with these regions appearing to increase in size as the host ARs aged. No systematic changes in the frequency of IRIS burst profiles or the spectral characteristic of IRIS burst profiles through time were found for these ARs. Finally, $93$ \%\ of parent IRIS bursts with areas between $1$ arcsec$^2$ and $4$ arcsec$^2$ were observed to occur co-spatial to bi-poles in the photosphere.}
{IRIS bursts are small-scale brightenings which have remarkably consistent spectral and spatial properties throughout the evolution of ARs. These events predominantly form within the cores of larger and more complex ARs, with the regions containing these events appearing to increase in size as the host region itself evolves.}

\authorrunning{Nelson et al.}
\titlerunning{IRIS Burst Properties}

\maketitle

\section{Introduction}
\label{Introduction}

Small-scale burst activity has been widely observed across the broad temperature range of the solar atmosphere, from Ellerman bombs (EBs) in the photosphere (see, for example, \citealt{Ellerman17, Vissers13, Nelson15}), through UV bursts in the transition region (as discussed by \citealt{Peter14, Judge15, Young18}), to the recently discovered campfires in the corona (e.g. \citealt{Berghmans21, Mandal21, Panesar21}). Each of these dynamic events is thought to be driven by localised magnetic reconnection between opposite polarity magnetic fields, with the properties of the reconnection (e.g. height in the atmosphere, amount of energy released) seemingly dictating which type of burst is formed. Recent numerical simulations have supported this assertion (see: \citealt{Danilovic17} for EBs; \citealt{Hansteen17} for UV bursts; and \citealt{Chen21} for campfires) providing us with a clearer understanding of the physical mechanisms responsible for these events. Despite this general understanding, there is still much we don't know about these features, including whether UV bursts are detected in all active regions (ARs) and whether their properties (potentially diagnosing the physics of the magnetic reconnection driving these events) change during the evolution of the host ARs. 

A new sub-set of UV bursts, characterised by large increases in intensity in spectral lines sampling the transition region (including the \ion{Si}{IV} $1394$ \AA\ and $1403$ \AA\ lines), were identified recently in data from the Interface Region Imaging Spectrograph (IRIS; \citealt{DePontieu14}) by \citet{Peter14}. Those authors studied a raster sampling AR $11850$ and found four burst events, which were later named IRIS bursts in the literature, co-spatial to bi-poles in the photosphere. It was suggested that these events could be driven by magnetic reconnection heating the local plasma to temperatures close to $80$ kK. The presence of several chromospheric absorption lines (e.g. \ion{Ni}{I} $1393$ \AA) over-laid on the IRIS burst spectra, however, indicated that these events occurred deep in the solar atmosphere, potentially at heights as low as several hundred km. Subsequent research found that IRIS bursts could occur co-spatial to EBs (\citealt{Vissers15, Tian16}), which are known to have temperatures well below the typical formation temperatures of the \ion{Si}{IV} lines (see, for example, \citealt{Fang06, Hong14}), and cancelling bi-poles (e.g. \citealt{Nelson16}) in the photosphere. Dedicated 1D modelling efforts have so far been unable to reproduce co-spatial EB and IRIS burst spectra, in part due to their large temperature differences (e.g. \citealt{Reid17, Hong17}). 

In order to better assess the overall impact of IRIS bursts on the global solar atmosphere, \citet{Kleint22} conducted an extensive statistical analysis of IRIS burst profiles using the more than $3500$ IRIS spectral datasets sampled during 2013 and 2014 (utilising the methods detailed in \citealt{Panos21a} and \citealt{Panos21b}). Those authors considered a \ion{Si}{IV} $1394$ \AA\ spectra to be IRIS burst-like if absorption lines were over-laid on it, agreeing with one of the key observational characteristics of these events as noted by \citet{Young18}. More than $100$ k IRIS spectra satisfied this criteria (within $287$ datasets), indicating that around $0.01$ \% of all \ion{Si}{IV} $1394$ \AA\ spectra recorded by IRIS are IRIS burst profiles. Of the sample reported, no IRIS burst profiles were detected in the quiet-Sun or close to the solar poles suggesting that these events occur exclusively in ARs. It is likely, therefore, that the potential quiet-Sun IRIS burst presented by \citet{Nelson17} would not have been classed as being IRIS burst-like by this algorithm (if it had been applied to that dataset) due to the lack of clear absorption lines over-laid on the relatively narrow \ion{Si}{IV} $1394$ \AA\ line. Finally, no systematic increase in emission co-spatial to IRIS burst profiles was found either for the IRIS \ion{Fe}{XXI} line or for the EUV lines sampled by the Solar Dynamics Observatory's Atmospheric Imaging Assembly (SDO/AIA; \citealt{Lemen12}) by \citet{Kleint22} suggesting that the magnetic reconnection hypothesised to be driving these events does not heat the local plasma to coronal (>10$^6$ K) temperatures.

Although we focus specifically on IRIS bursts in this article, it should be noted that a whole host of other burst events have also been detected in the solar atmosphere in spectral windows sampling similar temperature ranges. This wider family of events are collectively known as UV bursts, and include features such as explosive events (\citealt{Dere89}) and blinkers (\citealt{Harrison97}). Whilst one of the unique defining properties of IRIS bursts are the characteristic overlying absorption lines on the \ion{Si}{IV} $1394$ \AA\ spectra, other types of UV bursts have a plethora of other spectral shapes (some are extremely narrow whilst some are extremely broad) and properties (some are symmetric about the spectral line whilst some display broadenings in only the red or blue wings). This large variance in spectral signatures means other UV burst types will not be identified by the version of the algorithm developed by \citet{Kleint22}. The inclusion of more template spectra would be required if these other types of UV bursts were to be studied in an automated manner in the future. The spectral differences between distinct types of UV bursts also means that the defining properties of this family of events must be taken from other observable properties, with key signatures being that they are: i) compact (areas typically below $2$\arcsec); ii) short-lived (lifetimes of up to around one hour); iii) intense compared to the local background; iv) relatively stationary; and v) not linked to larger-scale flares. Given these properties it is, therefore, possible for an event to display an IRIS burst profile but not be a UV burst. For a comprehensive review of both IRIS bursts and the wider family of UV bursts see \citet{Young18}.

The algorithm developed by \citet{Kleint22} provides numerous opportunities for advancing our understanding of the importance of IRIS bursts in the solar atmosphere. In this article, we utilise a slightly modified version of this algorithm to investigate how the measurable properties of IRIS bursts (e.g. frequency, area, location, spectral characteristics) in large IRIS rasters vary during the evolution of seven ARs. The motivation behind this research is to identify whether there are preferential conditions for when and where magnetic energy is released in ARs as they evolve. We structure our work as follows: In Sect~\ref{Methods} we detail the data products analysed and outline the changes made to the IRIS burst detection algorithm; In Sect.~\ref{Results} we present the results of our research and a brief discussion; before in Sect.~\ref{Conclusions} we draw our conclusions.

\section{Methods}
\label{Methods}

\subsection{Observations}

The primary data products analysed in this research were sampled by IRIS, a highly versatile instrument which can be employed in a plethora of observing modes allowing the user to vary the spatial, temporal, and spectral resolutions of the returned data based on scientific need. This versatility creates a comprehensive data catalogue but also means we must filter the available datasets in order to isolate only those of interest for this particular study. Here, we focus our analysis on ARs that were sampled by a minimum of four separate $400$-step IRIS rasters in the time-period studied by \citet{Kleint22}. These observations must include the \ion{Si}{IV} $1394$ \AA\ line, and have time differences between the first and final rasters of at least 48 hours. Seven ARs (11850, 11856, 11871, 11909, 11916, 12104, and 12139) satisfied these criteria, forming the sample we analyse here. Of these, two ARs (11871 and 11916) emerged onto the solar disk between one and three days prior to their first IRIS raster scans, whilst the remaining five (11850, 11856, 11909, 12104, and 12139) rotated onto the solar disk as fully formed ARs between four and six days prior to the first IRIS raster scans which sampled them.

\begin{figure}
\center
\includegraphics[width=0.49\textwidth]{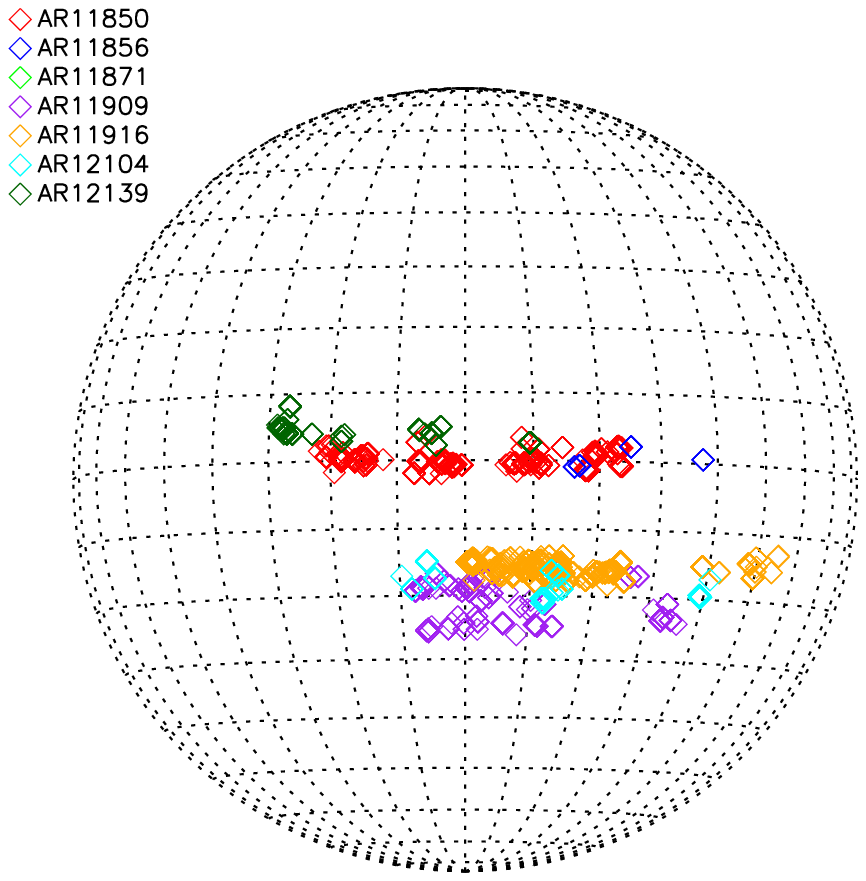}
\caption{Locations on the Sun of each of the IRIS burst profiles studied in this article, with different colours corresponding to different ARs. IRIS burst profiles are detected in multiple rasters sampled over multiple days for six of the seven ARs. Clearly, no IRIS burst profiles were detected at any time within AR 11871. General information about the identified IRIS burst profiles is presented in Table~\ref{Tab_overview}.}
\label{Fig_locations}
\end{figure}

\begin{figure*}
\center
\includegraphics[width=0.5\textwidth,trim={0.3cm 0.5cm 0 0.1cm,clip}]{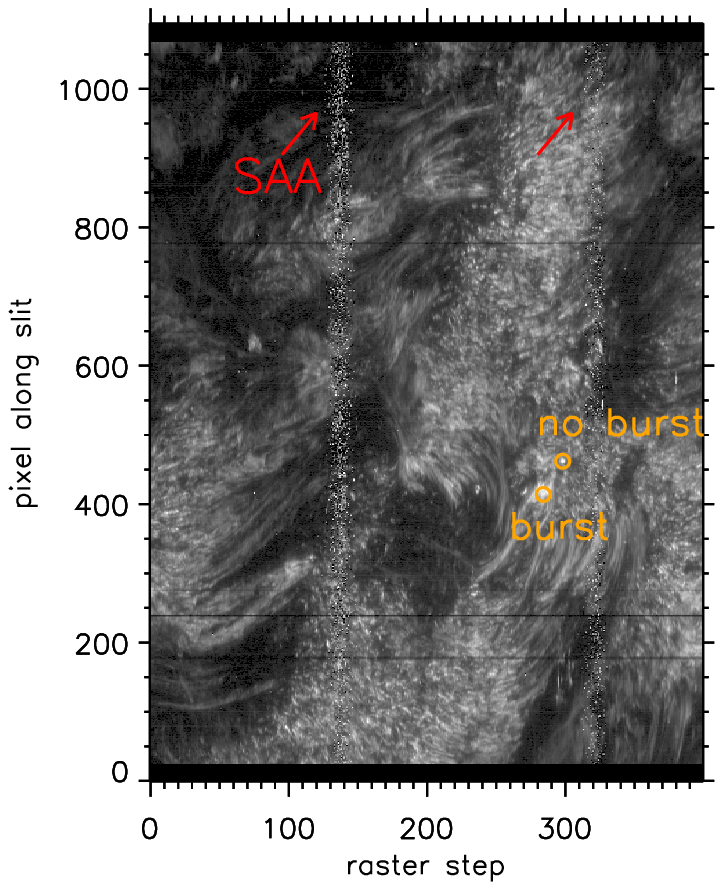}
\includegraphics[width=0.45\textwidth]{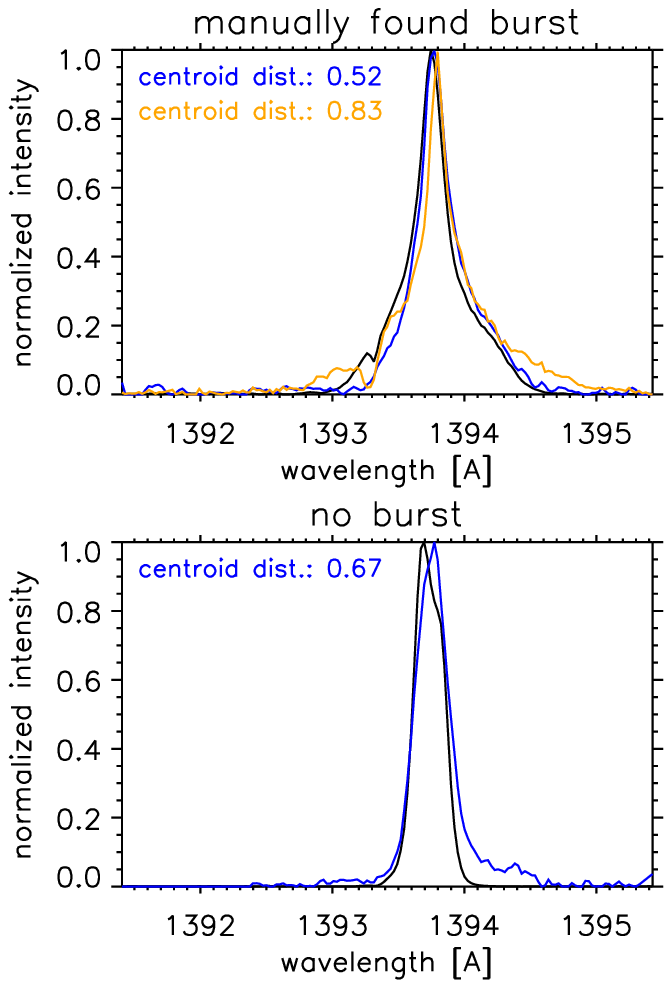}
\caption{Example of IRIS burst profiles missed by the algorithm. (Left-hand panel) Intensity of the FOV during the fourth IRIS raster sampling AR $12104$ at the core wavelength of the \ion{Si}{IV} 1394 line (logarithmically scaled to enhance contrast). Our algorithm did not identify any bursts in this raster. The two times during which IRIS passed through the SAA, thereby causing noisy spectra, during this observation are marked by red arrows. The orange circles locate two pixels whose spectra are shown in the right-hand panels (black lines), one containing a manually detected burst and the other containing no burst spectra. (Right-hand panels) The two IRIS spectra selected to demonstrate how a small number of IRIS burst profiles may be missed by the algorithm. The top panel plots a manually detected IRIS burst with a clear, albeit shallow, absorption blend over-laid on the \ion{Si}{IV} $1394$ \AA\ line, while the bottom panel is a typical non-burst spectrum. Both spectra are normalised to their peak intensity. We over-plot the mathematically closest reference spectrum considered by the algorithm to each of these profiles (blue lines), which in both cases is a non-burst reference spectrum. Therefore, no IRIS burst profile is found for either pixel by the algorithm. The nearest IRIS burst reference spectrum to the manually detected IRIS burst profile is also over-plotted on the top panel for reference (orange line).}
\label{Fig_manual}
\end{figure*}

Overall, $42$ IRIS datasets were analysed in this study. Each of these datasets had a raster step size of $0.35$\arcsec\ and a pixel scale along the slit of $0.17$\arcsec, allowing us to consistently measure the properties of IRIS bursts. General information about each of these datasets is provided for reference in Table~\ref{Tab_overview}. From this sample, $35$ datasets were found to contain a combined total of $4019$ IRIS burst profiles using the modified IRIS burst detection algorithm. This accounts for $0.02$ \%\ of the total number of IRIS \ion{Si}{IV} $1394$ \AA\ profiles analysed here. To supplement this automated detection, we also conducted a manual analysis of the remaining seven datasets using CRISPEX (\citealt{Vissers12}) in order to identify whether any clear and obvious IRIS burst profiles were missed by the algorithm. IRIS burst profiles, identified through the presence of absorption lines on the \ion{Si}{IV} $1394$ \AA\ spectra for consistency, were detected in one further dataset through these checks. The remaining six datasets did not appear to contain any clear and obvious IRIS burst profiles (we cannot definitively state that a small number of pixels did not evade our manual search which will have been biased towards more intense pixels). The locations on the Sun of each of the IRIS burst profiles detected using automated methods are plotted in Fig.~\ref{Fig_locations}. The different coloured boxes correspond to different ARs, as detailed in the legend.

Given longer exposure times may lead to an increase in the number of saturated pixels (which are removed from the sample before the algorithm is applied), it is also important to investigate the numbers of potential IRIS burst profiles removed from each of these datasets before any additional analysis is conducted. In order to test this, we identified all pixels which were saturated (counts$>16180$) at one or more spectral positions in the wavelength range $1393.75\pm0.2$ \AA. In total $239$ pixels matched this criterion across all $42$ datasets with approximately $100$ of these appearing to be caused by cosmic ray spikes and another $100$ appearing to be some other kind of burst, with relatively narrow spectra and no absorption lines over-laid on the \ion{Si}{IV} $1394$ \AA\ spectra. This, therefore, left around $40$ potential IRIS burst profiles which were spread across four datasets. Around half of these saturated potential IRIS burst profiles were found in the first dataset sampling AR $11916$, which already contained $715$ IRIS burst profiles, whilst the others were detected in the seventh raster to sample AR $11916$, the first raster sampling AR $12104$, and the third raster sampling AR $12104$. Overall, the small numbers of saturated pixels displaying potential IRIS burst profiles (less than $1$ \%\ of the sample studied here) and the centralisation of these to only a small number of datasets indicates that the removal of saturated pixels does not overly influence the results obtained in this manuscript.

We also made use of data from SDO's Helioseismic and Magnetic Imager (SDO/HMI; \citealt{Scherrer12}) in order to study the line-of-sight (LOS) magnetic field co-spatial to the IRIS observations. Full-disk LOS magnetic field maps were downloaded with a cadence of one hour in the time-period spanning from six hours prior to the start time of the first IRIS raster to six hours after the end time of the last IRIS raster for each AR. A zoomed $180$\arcsec$\times180$\arcsec\ field-of-view (FOV) cut-out was then created co-temporal to the first IRIS scan centred on the pointing co-ordinates of that raster. All other SDO/HMI frames were then de-rotated to that time, before a co-spatial cut-out was constructed in order to create a consistent FOV through the entire time-period. Comparisons of these SDO/HMI maps with plage regions in the IRIS rasters indicated that these alignments were reasonable (estimated errors $<5$\arcsec). As we only study the overall behaviour of the magnetic field within the entire ARs and the general structure of the magnetic field co-spatial to the IRIS burst profiles through time, no further alignment between instruments was conducted. These SDO/HMI data have a post-reduction pixel scale of $0.6$\arcsec\ which is more than adequate for investigating the overall behaviour of these ARs.

\subsection{Modifications to the IRIS burst detection algorithm}
\label{Obs_mod}

To detect IRIS burst profiles, the algorithm of \citet{Kleint22} was used with slight modifications applied in order to both be able to study rasters that may have a small number of IRIS burst profiles and fix two minor bugs, which became apparent during this analysis. To tackle the first issue, the criterion of requiring at least 30 potential IRIS burst profiles applied by \citet{Kleint22} was dropped and any observation that had at least one candidate IRIS burst profile was analysed further. This increased the sample of IRIS burst profiles detected within these $42$ rasters from $2462$ to $3113$ (a $26.44$ \%\ increase). Following this adaption, the first bug, related to times when the IRIS orbit passes through the South Atlantic Anomaly (SAA) and hence returns noisy spectra, became apparent in the results. Specifically, the IRIS FITS SAA level2 keyword defaults to the value that is true for \textit{most} of the observation. Therefore, observations where more time was spent in the SAA than outside of the SAA (which is very rare) will have this keyword set to 1 at all times and were, thus, excluded in the first version of the pipeline. The first version of the pipeline correctly queried the level1 data, which includes the actual flag for SAA/no SAA and updated the corresponding header entry in the level2 data to 1 during times of SAA. However, it did not check if the header entry may already be set to 1 due to the convention of the level2 data for SAA dominant datasets, even though no SAA occurred at the specific time when the spectra were sampled. The new version of the pipeline does both checks. 

The second bug is related to the fact that the \ion{Si}{iv} observing window width changed slightly with the evolution of the IRIS linelists, and between the different types of linelists. The original pipeline had a bug with extrapolations of spectra, leading to unintended values far away from the \ion{Si}{iv} lines. This did not affect strong bursts (since the signal is near the line center and the continuum is close to zero due to the normalization), which have always been classified properly, but it could mistakenly classify noisy burst spectra into a non-burst category. Extrapolated values are now properly set to zero and thus do not affect the classification. The classification in \citet{Kleint22} focused on minimizing false-positives and if in doubt, a spectrum was excluded from being of burst-type so this should not have influenced their results excessively. Applying the algorithm following these subsequent modifications increased the sample further from $3113$ IRIS burst profiles to a total of $4019$ IRIS burst profiles (a further increase of $29.10$ \%). These new IRIS burst profiles mostly have very weak absorption lines and/or are noisy, but do visually look like bursts so are included here for completeness. Notably, these events were detected across a total of $35$ datasets, up from $15$ using the algorithm before removal of the $30$ spectra criterion and the subsequent modifications.

\begin{figure*}
\center
\includegraphics[width=0.99\textwidth,trim={0 2cm 0 0}]{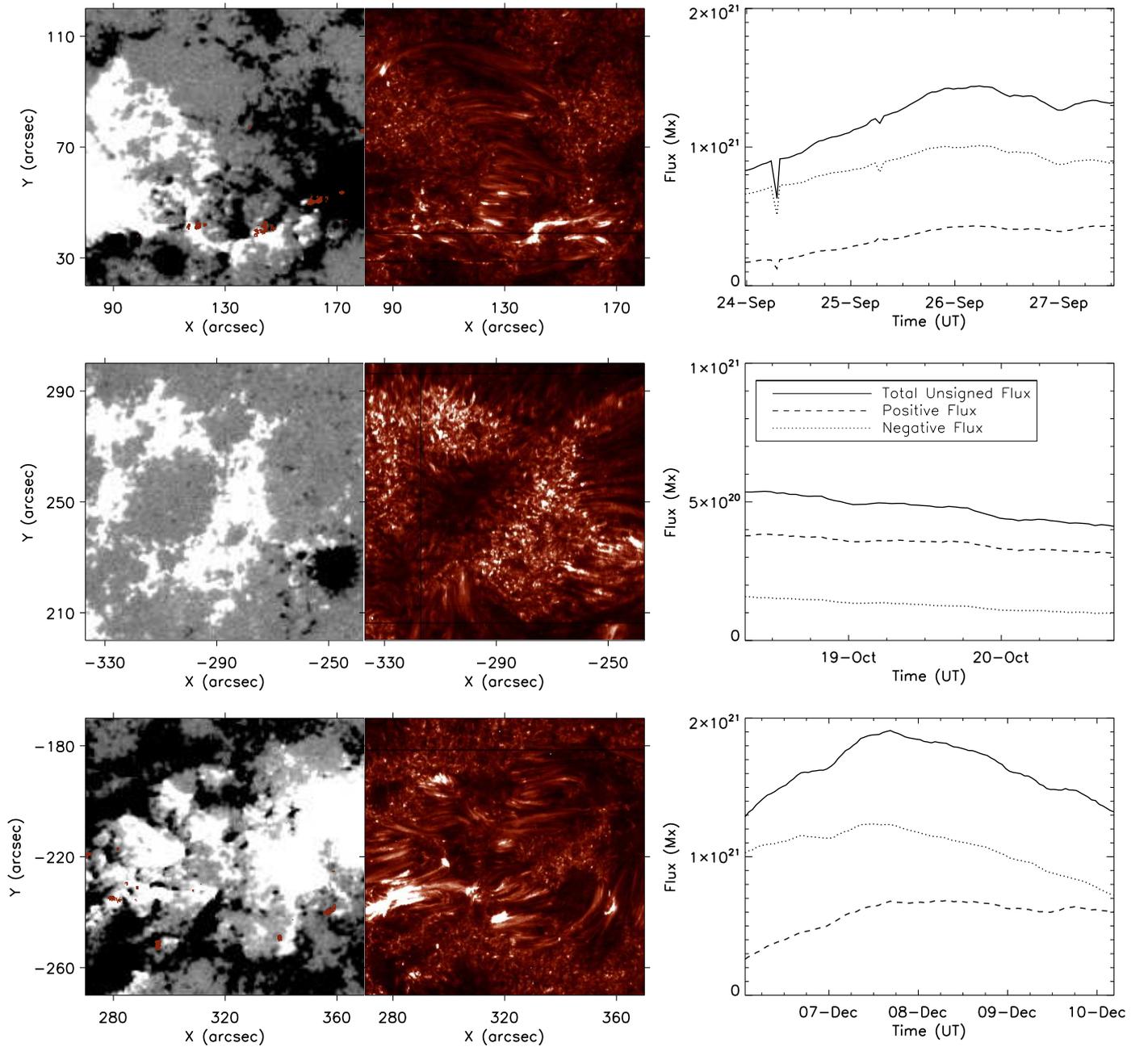}
\caption{Context plots for three of the ARs, namely ARs 11850, 11871, and 11916 from top to bottom, studied. (Left-hand column) Zoomed $100$\arcsec$\times100$\arcsec\ SDO/HMI magnetic context images. The plotted frames were selected as they contained the peak total unsigned magnetic flux for these regions during the time studied. (Middle column) The \ion{Si}{IV} $1394$ \AA\ line core intensity sampled from the IRIS raster with start time closest to the SDO/HMI images plotted in the left hand column. The pointing of the IRIS FOV was de-rotated to the SDO/HMI observation time. (Right-hand column) The evolution of the total unsigned (solid lines), positive (dashed lines), and negative (dotted lines) magnetic flux within the wider $180$\arcsec$\times180$\arcsec\ FOV through the time-period studied here for each AR. The red contours over-laid on the left-hand column outline the IRIS burst profiles returned using automated methods from the IRIS raster plotted in the respective middle column. The five ARs with the highest peak magnitudes of the unsigned flux were found to contain the majority of the IRIS burst profiles identified here.}
\label{Fig_context}
\end{figure*}

\begin{figure*}
\center
\includegraphics[width=0.99\textwidth]{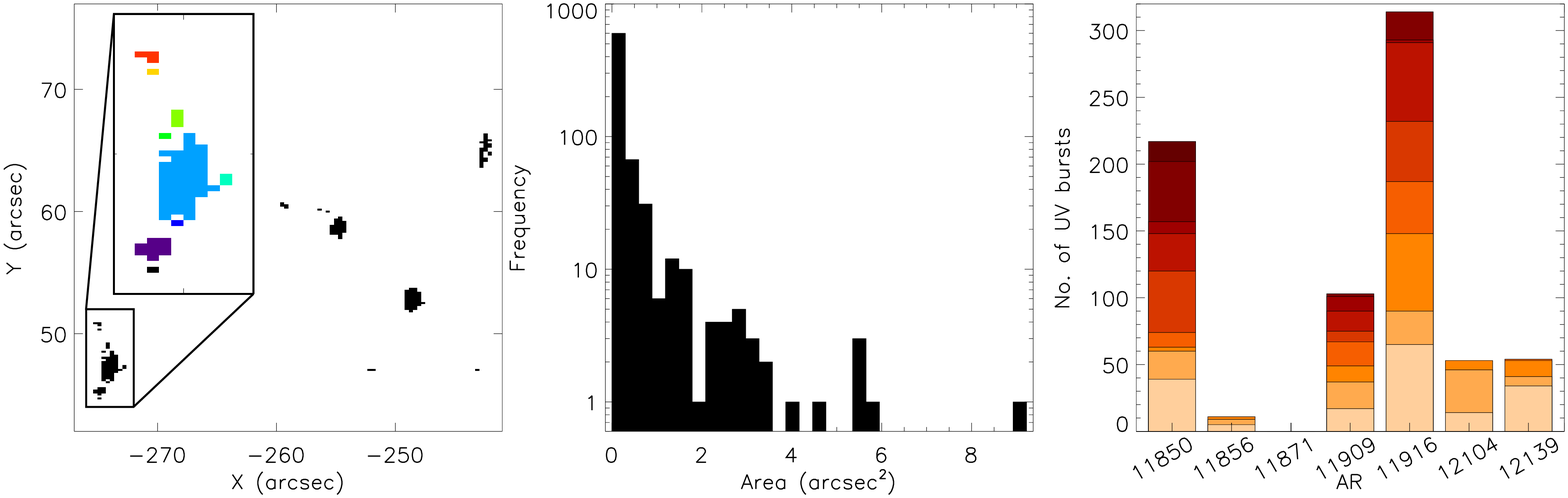}
\caption{Structuring of IRIS burst profiles in these datasets. (Left-hand panel) Binary map showing the locations of IRIS burst profiles (black pixels) identified within a single raster sampling AR 11850. Four of the parent IRIS bursts within this raster correspond to the four events studied in \citet{Peter14}. The over-lying cut-out displays how connected objects are combined to form parent IRIS bursts, with nine events (all individually coloured) occurring in this small region. (Middle panel) Histogram of parent IRIS burst area against frequency constructed using all $42$ datasets, where the $y$-axis is plotted in logarithmic scale. Each bar is binned over a range of $0.3$ arcsec$^2$ (corresponding to five pixels). (Right-hand panel) Bar chart plotting the number of parent IRIS bursts by AR. The different coloured sections of each bar indicate the number of parent IRIS bursts identified in different rasters sampling that AR, with the light yellow at the bottom corresponding to the first raster within which IRIS bursts were detected using automated methods for that specific AR. The progressively darker colours up the bars correspond to the subsequent rasters within which IRIS bursts were detected. The physical number of bursts plotted in each bar segment can be found in Table~\ref{Tab_overview}.}
\label{Fig_stats}
\end{figure*}

It should be mentioned, however, that the variety of burst spectra makes it impossible to catch every single burst with a classification template of ``only'' 423 different spectral types. We are confident, though, that the most common types of burst spectra are identified reliably. An example of an IRIS burst profile which was not detected by the pipeline is displayed in Fig.~\ref{Fig_manual}. In the left-hand panel we plot the intensity of the \ion{Si}{IV} $1394$ \AA\ line core for the fourth raster sampling AR 12104 (i.e. the raster where IRIS bursts were detected manually but not automatically). The locations where IRIS passed through the SAA during this observation are marked by the red arrows. The orange circles outline two pixels, one containing a manually detected IRIS burst profile and one containing a non-burst profile, chosen for reference. In the right-hand panels, the black lines plot the observed spectra at each of these pixels with clear, albeit shallow, absorption blends indicative of an IRIS burst profile visible on the top panel. The blue lines plot the reference spectra deemed to be the closest mathematically to the observed spectra which in both cases is a non-burst reference spectra. The orange line included on the top panel plots the closest IRIS burst reference spectrum to this spectrum, which was mathematically `further' from the observed spectrum than the non-burst reference spectrum (distance of 0.83 compared to a distance of 0.52). The only way to improve detections of such bursts would be to increase the number of template reference spectra from the 423 currently used. Given we only found a very small number (around ten) of additional IRIS burst profiles through our manual checks, and adding further reference spectra risks increasing the rate of false detections, we did not deem this necessary for this study.

\section{Results and discussion}
\label{Results}

\subsection{General AR properties}

To begin our analysis, we investigated the general properties of each AR in order to better contextualise any subsequent inferences about IRIS bursts. Firstly, we identified the number of raster scans which displayed IRIS burst profiles for each AR finding varying behaviour. Of the seven ARs studied here, five (11850, 11909, 11916, 12104, and 12139) contained IRIS burst profiles in each raster scan, including both IRIS burst profiles detected using automated methods and those identified through the subsequent manual analysis. AR 11916 contained the most automatically detected IRIS burst profiles, with $2052$ ($51.06$ \%\ of the total sample studied here) being identified over the eight rasters analysed, whilst the other four ARs each contained between $1144$ and $168$ automatically detected IRIS burst profiles. On top of this, one AR (11856) was found to contain IRIS burst profiles in some but not all rasters and one AR (11871) was found to contain no IRIS burst profiles at any time. This analysis indicates that the age of the AR is not a defining factor in whether IRIS burst profiles are present with the two youngest ARs (11871 and 11916), which both emerged onto the solar disk in the three days prior to the IRIS scans, displaying sharply contrasting behaviour.

Secondly, we calculated the total unsigned magnetic flux within a zoomed $180$\arcsec$\times180$\arcsec\ FOV around each AR at each SDO/HMI time-step. Interestingly, the five ARs which contained IRIS burst profiles in each raster had the five highest peak fluxes ($1.44\times10^{21}$ Mx, $1.66\times10^{21}$ Mx, $1.91\times10^{21}$ Mx, $2.23\times10^{21}$ Mx, and $1.50\times10^{21}$ Mx for ARs 11850, 11909, 11916, 12104, and 12139, respectively) with each exceeding $10^{21}$ Mx. The remaining two ARs, which contained either a small number or zero IRIS burst profiles, both had peak fluxes of less than $10^{21}$ Mx ($8.60\times10^{20}$ Mx and $5.38\times10^{20}$ Mx for ARs 11856 and 11871, respectively). In the left-hand column of Fig.~\ref{Fig_context}, we plot $100$\arcsec$\times100$\arcsec\ zoomed SDO/HMI images sampled at the time at which the peak unsigned magnetic flux was measured for three ARs (11850, 11871, and 11916 from top to bottom, respectively). The middle column of Fig.~\ref{Fig_context} plots the IRIS \ion{Si}{IV} $1394$ \AA\ line core intensity across this FOV sampled at the closest raster to the SDO/HMI images, with the pointing corrected for solar rotation. The red contours over-laid on the SDO/HMI images outline the locations of the IRIS burst profiles identified in these rasters using automated methods. The right-hand column of Fig.~\ref{Fig_context} plots the evolution of the total unsigned magnetic flux (solid line), positive flux (dashed line), and negative flux (dotted line) calculated from the larger $180$\arcsec$\times180$\arcsec\ SDO/HMI FOV with time. Clearly, ARs 11916 and 11850 are more complex in their magnetic structuring (left-panels of Fig.~\ref{Fig_context}) and have total unsigned magnetic fluxes which initially increase before reaching a peak and then decreasing (right-panels of Fig.~\ref{Fig_context}). On the other hand, AR 11871 is less complex and has a total unsigned magnetic flux which slowly decreases with time. Our results indicate that the total unsigned magnetic flux and the complexity of the structuring of that flux within an AR may be a dominant factor determining whether IRIS bursts are widely detected or not, rather than the age of the AR.

\subsection{Spatial properties of IRIS burst profiles}

\begin{figure*}
\center
\includegraphics[width=0.99\textwidth]{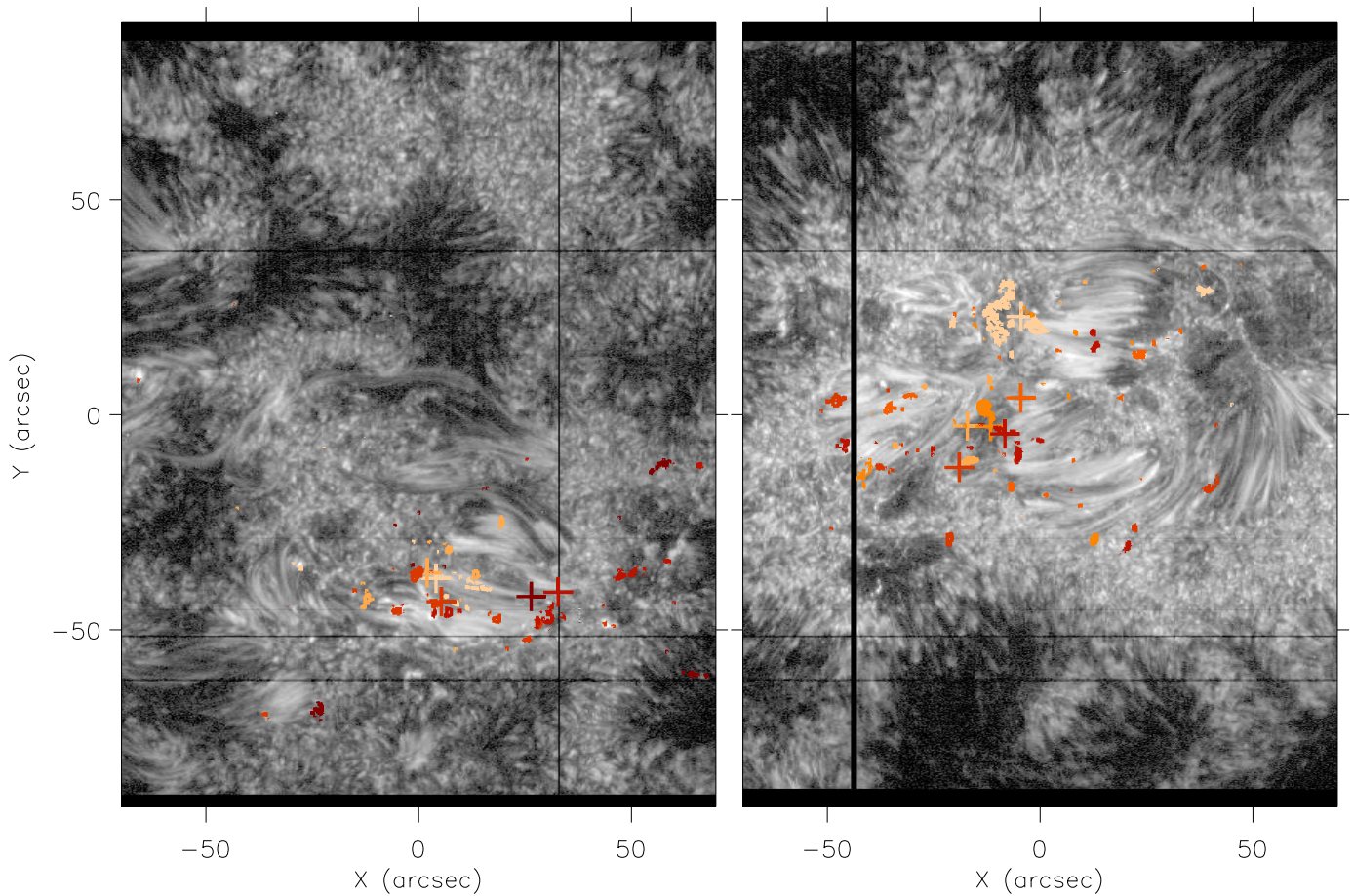}
\caption{Spatial positions of IRIS burst profiles through time in two datasets. (Left-hand panel) Map of the \ion{Si}{IV} $1394$ \AA\ line core intensity (logarithmically scaled) within AR 11850 sampled during the raster starting at 05:59:43 UT on 26th September 2013. (Right-hand panel) Same as the left-hand panel but for AR 11916 sampled during the raster starting at 03:09:49 UT on 7th December 2013. The contours and crosses over-laid on both panels denote the IRIS burst profiles and their centres-of-mass, respectively, for each raster containing more than 100 automatically detected IRIS burst profiles. The colours indicate different rasters and are consistent with those used in the right-hand panel of Fig.~\ref{Fig_stats}.}
\label{Fig_loc}
\end{figure*}

We continued our analysis by grouping the $4019$ IRIS burst profiles identified here into separate, connected objects which we define as parent IRIS bursts. To achieve this, we created binary maps of the locations of IRIS burst profiles within each raster scan and then combined IRIS burst profiles which were adjacent in the $x-y$ plane together into parent IRIS bursts. If two parent IRIS bursts were separated by even one pixel, we considered that they were different events. Additionally, if the observational routine had more than one raster then only the middle raster was considered here in order to avoid longer-lived parent IRIS bursts, potentially present in repeated rasters, being counted multiple times. Importantly, our analysis of these parent UV bursts does not include any temporal component. This stipulation only removed $119$ IRIS burst profiles from our sample meaning it should not have a major influence on our results. Overall, this analysis returned $752$ parent IRIS bursts spread across the $35$ IRIS rasters. In the left-hand panel of Fig.~\ref{Fig_stats}, we plot an example of a binary map displaying the locations of numerous IRIS burst profiles from one specific raster within AR 11850. This plot is calculated from the dataset which was used by \citet{Peter14} to originally identify IRIS bursts, with the four largest parent IRIS bursts corresponding to the four events studied by those authors. This clearly shows the effectiveness of the algorithm developed by \citet{Kleint22} and modified here. The zoomed cut-out over-laid on this panel more clearly displays how adjacent IRIS burst profiles are combined to form separate parent IRIS bursts. Nine parent IRIS bursts are present within this small FOV, with each event being plotted in a different colour.

In the middle panel of Fig.~\ref{Fig_stats}, we plot a histogram of parent IRIS burst area against frequency for all of the $752$ identified events. The histogram is binned across $0.3$ arcsec$^2$ (five pixels) and the $y$-axis is plotted using a logarithmic scale. Clearly, the majority of the studied events are small, with $383$ of these parent IRIS bursts having areas of only one pixel. The mean area of this sample of parent IRIS bursts is $0.31$ arcsec$^2$ (corresponding to 5.19 pixels); however, the distribution is non-Gaussian and the standard deviation of this sample is $0.72$ arcsec$^2$ (12.18 pixels), more than double the mean, suggesting one should not interpret this as a precise estimate. Our results do, though, clearly indicate that the majority of parent IRIS bursts are extremely small-scale. Notably, several larger brightenings in the \ion{Si}{IV} $1394$ \AA\ line core are found to contain numerous parent IRIS bursts through our analysis. For example, one $6.30$\arcsec$\times2.72$\arcsec\ region in the second raster sampling AR 12104 (around $x_\mathrm{c}\approx200$\arcsec, $y_\mathrm{c}\approx-296$\arcsec) was found to contain $20$ parent IRIS bursts, with an average area of $0.12$ arcsec$^2$ (corresponding to 1.57 pixels). From the entire sample studied here, only 52 parent IRIS bursts (6.91 \%) have an area over $1$ arcsec$^2$ (corresponding to 17 pixels) with 28 of these occurring in AR 11916. The largest parent IRIS burst occurred in AR 11916 and had an area of $8.98$ arcsec$^2$ (corresponding to $151$ pixels), much larger than the typical sizes of UV bursts in general (see \citealt{Young18}). 

\begin{figure*}
\center
\includegraphics[width=0.99\textwidth]{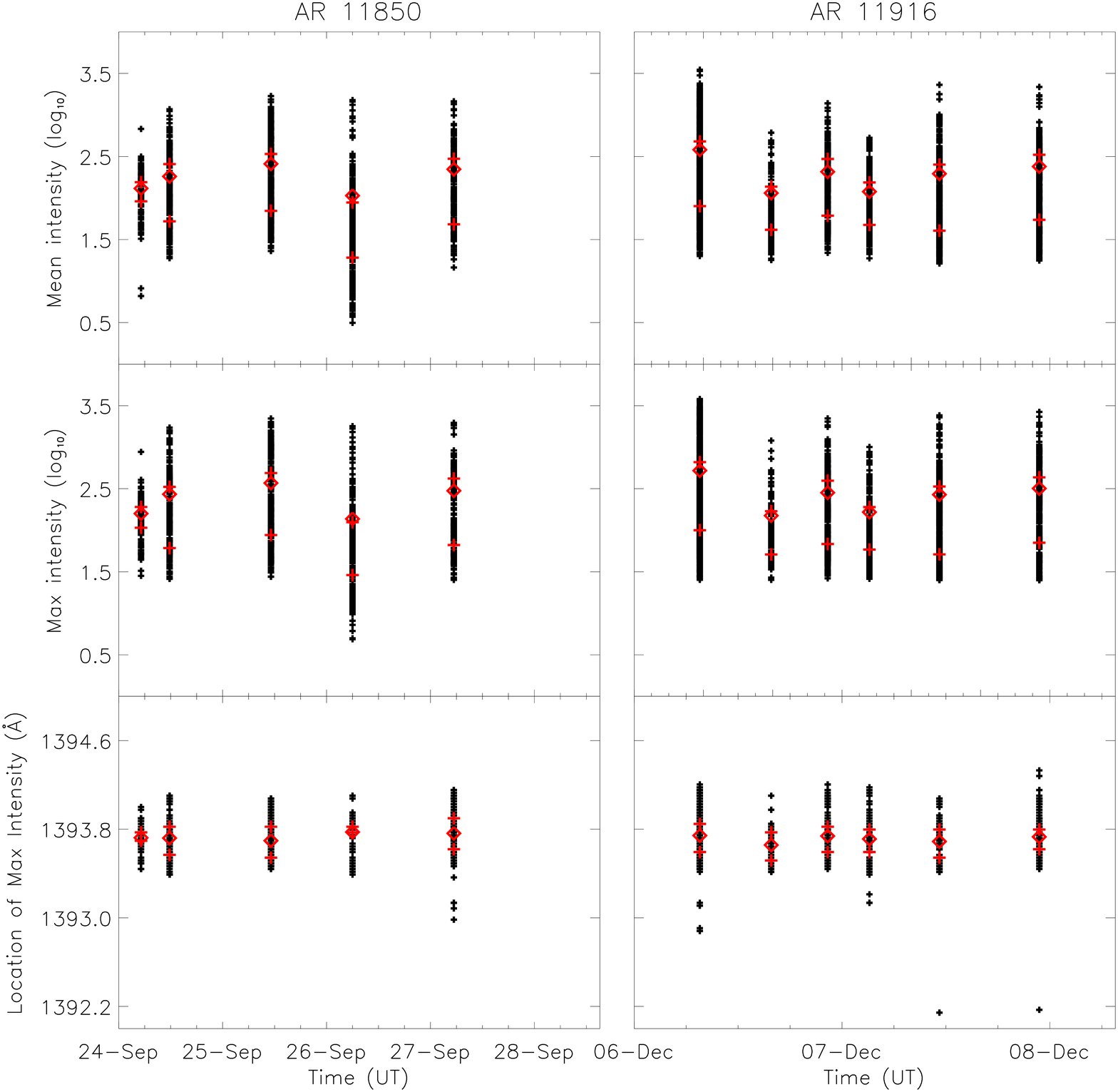}
\caption{Distributions of various spectral parameters for all IRIS burst profiles identified within datasets sampling ARs 11850 (left-hand panels) and 11916 (right-hand panels). Only datasets which returned more than $100$ IRIS burst profiles were considered. (Top panels) Logarithmic mean intensity from a $\pm0.075$ \AA\ window surrounding the \ion{Si}{IV} $1394$ \AA\ line core calculated for each IRIS burst profile. (Middle panels) Logarithmic maximum intensity for each IRIS burst profile. (Bottom panels) Spectral location of the maximum intensity of each IRIS burst profile. The diamonds indicate the mean for each raster, while the upper and lower crosses indicate the $75$th and $25$th percentiles, respectively. No systematic or monotonic changes in these parameters through time are apparent.}
\label{Fig_spec}
\end{figure*}

In the right-hand panel of Fig.~\ref{Fig_stats}, we plot a bar chart displaying the number of parent IRIS bursts identified per AR. The colours in each bar indicate parent IRIS bursts identified in different rasters, with the bottom (light yellow) portion corresponding to the first raster within which IRIS burst profiles were identified for that specific AR. A high proportion of the parent IRIS bursts were found within ARs 11850 and 11916 ($70.61$ \% of the total sample) whilst AR 11871 contains zero parent IRIS bursts. The highest number of parent IRIS bursts in a single raster was 65, which were identified in the first raster sampling AR 11916, with the mean number of parent IRIS bursts per raster being 17.90. The average area of the parent IRIS bursts was found to change in a seemingly random way from raster to raster (neither monotonically increasing or decreasing) as the host ARs evolved. In total, ten rasters within the sample analysed here contained more than $30$ parent IRIS bursts, with five of these occurring in AR 11916, three occurring in AR $11850$, and one occurring in both ARs $12104$ and $12139$. Overall, these results imply a parent IRIS burst spatial frequency of $0.00069$ arcsec$^{-2}$ in ARs, well below the spatial frequency of, for example, spicules in the chromosphere. See Table~\ref{Tab_overview} for a summary of these findings.

\subsection{Locations of IRIS burst profiles within ARs}

One of the main aims of this research is to identify whether IRIS burst profiles are detected randomly across ARs or whether they are confined to consistent localised regions through time. In order to analyse this, we studied the locations of IRIS burst profiles within the two ARs (namely ARs 11850 and 11916) for which more than 100 IRIS burst profiles were detected using the algorithm in more than one raster. We de-rotated each raster to a consistent pointing which we converted as the origin (0\arcsec, 0\arcsec), constructed binary maps of the locations of IRIS burst profiles, and then calculated the centres-of-mass of IRIS burst profiles for each raster. In Fig.~\ref{Fig_loc}, we plot a visualisation of this analysis on maps of the \ion{Si}{IV} $1394$ \AA\ line core intensity (logarithmically scaled) from one representative raster for each AR (11850 in the left-hand panel and 11916 in the right-hand panel). The over-laid contours denote the locations of IRIS burst profiles and the crosses indicate the centres-of-mass for each specific raster. The relationship between colour and raster number is consistent with the colours used in the right-hand panel of Fig.~\ref{Fig_stats}.

\begin{figure*}
\center
\includegraphics[width=0.96\textwidth,trim={0.7cm 4.0cm 0 0}]{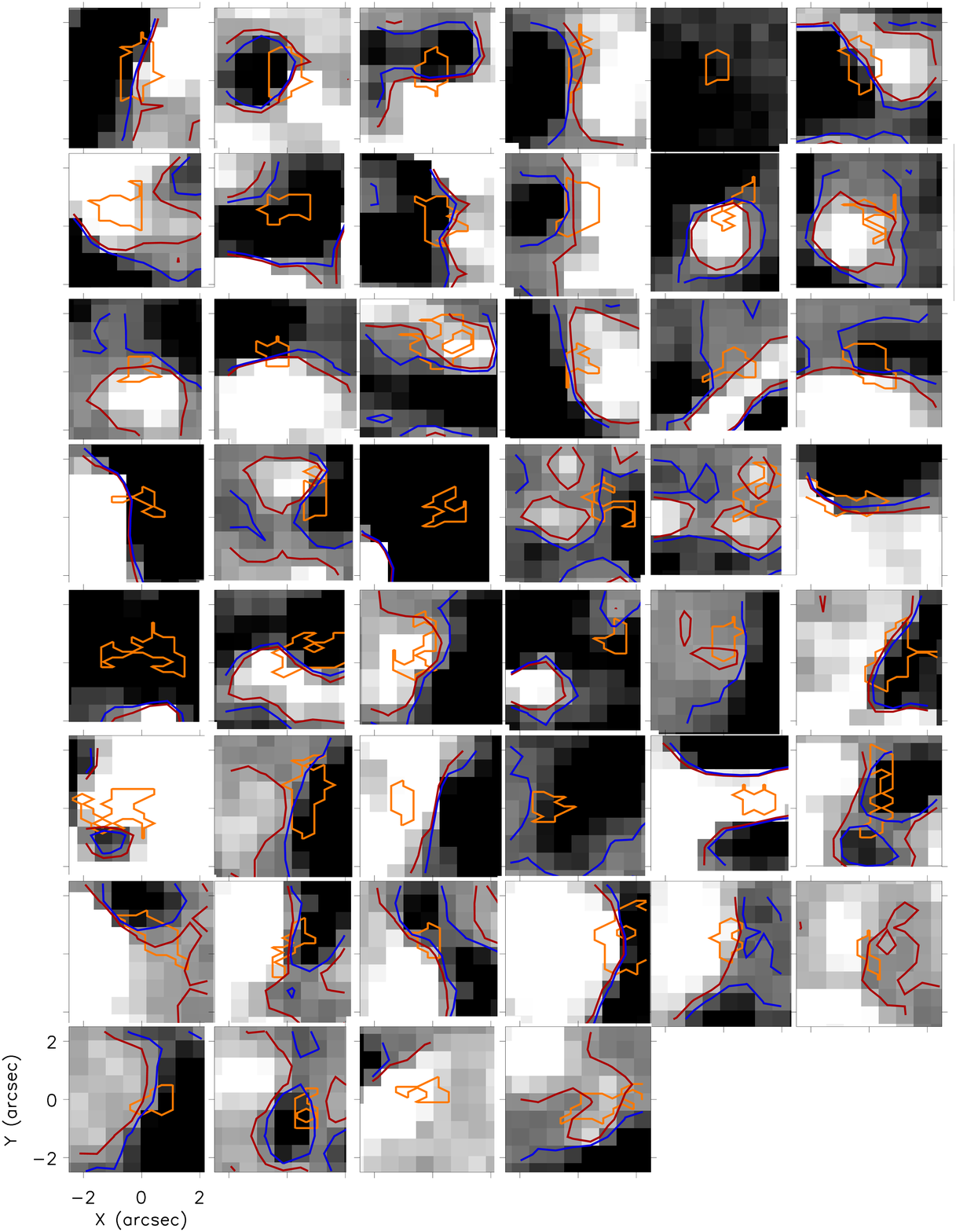}
\caption{LOS magnetic field maps co-spatial to the $46$ parent IRIS bursts with measured areas between $1$ arcsec$^2$ and $4$ arcsec$^2$. Each plot covers a region of $5$\arcsec$\times5$\arcsec\ centred on the location of the parent IRIS burst from the closest SDO/HMI image in time, de-rotated to the time of IRIS burst profile detection. The red and blue contours outline regions of positive and negative magnetic fields saturated at $20$ G and $-20$ G, respectively, whilst the orange contours outline the approximate locations of the parent IRIS bursts. Opposite polarity magnetic fields are found within the FOV for $43$ ($93$ \%) of these parent IRIS bursts.}
\label{Fig_mag}
\end{figure*}

For both ARs, IRIS burst profiles were found to be predominantly localised to relatively small regions, with approximate areas of $100$\arcsec$\times60$\arcsec. The centres-of-mass of IRIS burst profiles also appeared to be consistent through time for both ARs, only varying by a maximum of $31.11$\arcsec\ in AR 11850 and $37.96$\arcsec\ in AR 11916. In general, the locations of IRIS burst profiles within the ARs appeared to be limited to the most pronounced polarity inversion lines, as would be expected for events hypothesised to be driven by magnetic reconnection. Interestingly, IRIS burst profiles in both of these ARs appeared to become more spatially separated through time, with the contours in Fig.~\ref{Fig_loc} appearing further away from the centres-of-mass for subsequent rasters. In order to investigate this, we calculated the average separation between the IRIS burst profiles and the centres-of-mass through time for the rasters containing more than $100$ IRIS burst profiles. For AR 11850, this separation increased from $13.00$\arcsec\ ($\pm10.42$\arcsec) to $45.46$\arcsec\ ($\pm9.20$\arcsec) over the five rasters studied here; whilst for AR 11916, the average separation increased from $9.51$\arcsec\ ($\pm10.32$\arcsec) to $32.10$\arcsec\ ($\pm15.07$\arcsec) over the first five rasters before decreasing again to $18.63$\arcsec\ ($\pm15.04$\arcsec) by the sixth raster. Although this analysis indicates there may be an increase in the separation of IRIS burst profiles within an AR as it evolves, we note that two ARs were studied and large standard deviations were calculated, meaning this result is still tentative and requires further analysis using a larger sample of ARs in the future.

\subsection{Spectral properties of IRIS burst profiles}

To continue our analysis, we next investigated whether several spectral properties of IRIS burst profiles varied systematically during the evolution of ARs or whether they remained relatively consistent. Once again, only rasters within ARs $11850$ (left-hand panels) and $11916$ (right-hand panels) which contained more than $100$ IRIS burst profiles were studied in order to provide appropriate statistics. In Fig.~\ref{Fig_spec}, we plot the distributions of three parameters calculated from the \ion{Si}{IV} $1394$ \AA\ spectra sampled at all pixels automatically identified as IRIS burst profiles, for the appropriate rasters. The top panels plot the mean intensity (logarithmically scaled) from a $\pm0.075$ \AA\ window around the \ion{Si}{IV} $1394$ \AA\ line core, the middle panels plot the logarithmically scaled peak intensity of the \ion{Si}{IV} $1394$ \AA\ line for the identified IRIS burst profiles, whilst the bottom panels plot the spectral positions where the maximum intensity was measured for each IRIS burst profile. The diamonds on each distribution indicate the mean value calculated for that raster, and the top and bottom crosses plot the $75$th and $25$th percentile values, respectively. Overall, although some changes in the distributions are observed between specific rasters, no systematic changes (e.g. monotonic intensity increases, transitions from blue to red shifts) in these parameters is detected as the ARs age.

Following on from this, we also studied which reference spectra were returned the most frequently as the `closest' fits to the IRIS burst profiles identified in each of these rasters. This analysis allowed us to infer whether any changes in the common spectral shapes of IRIS burst profiles are present through time. In these datasets, no single reference spectrum was found to be the most accurate mathematical match to more than $7.34$ \%\ of all of the IRIS burst profiles within a single raster. Additionally, unique reference spectra were found to make up, on average across these datasets, $47.57$ \%\ of all of the observed spectra indicating that a huge variety of spectral shapes of IRIS burst profiles exist for each dataset. These two combined facts suggest that no preferential reference spectrum is present in any given raster, with the identified IRIS burst profiles being randomly spread between different reference spectra.

\subsection{Relations to the local LOS magnetic field}

Our final analysis focused on the relationship between the parent IRIS bursts identified in this study and the co-spatial under-lying LOS magnetic field structure as inferred by the SDO/HMI instrument. It is well known that IRIS bursts often occur co-spatial to regions of opposite polarity magnetic field (see, for example, \citealt{Young18}) where magnetic reconnection is thought to be possible in the lower solar atmosphere. As the SDO/HMI instrument has a spatial resolution of approximately $1$\arcsec, we only studied parent IRIS bursts with areas greater than $1$ arcsec$^2$ (corresponding to approximately 17 pixels). Additionally, we removed the parent IRIS bursts with areas above $4$ arcsec$^2$ (corresponding to 70 pixels) as the foot-points of those events would be difficult to define in a consistent manner. These stipulations left $46$ parent IRIS bursts for analysis. In Fig.~\ref{Fig_mag}, we plot the LOS magnetic field saturated at $\pm100$ G sampled for a $5$\arcsec$\times5$\arcsec\ FOV surrounding the centre of each parent IRIS burst. The red and blue contours outline positive and negative polarity LOS magnetic fields at values of $20$ G and $-20$ G, respectively, while the orange contours outline the locations of the parent IRIS bursts. The magnetic field map for each event was constructed by de-rotating the closest temporal frame in the respective one hour cadence SDO/HMI dataset to the time at which the parent IRIS burst was detected, meaning the locations of the bursts with respect to the magnetic field are only an approximation. Although this method will have some inherent errors on small-scales, the alignment appears to work well for the full FOV (as can be seen in Fig.~\ref{Fig_context}) giving us confidence that a closer alignment is not required for this general analysis.

Overall, $43$ ($93$ \%) of the panels plotted in Fig.~\ref{Fig_mag} contain bi-poles in the LOS magnetic field within the $5$\arcsec$\times5$\arcsec\ surrounding the parent IRIS bursts. For the three remaining events, one occurs co-spatial to an apparently uni-polar region of positive polarity flux and the other two occur co-spatial to apparently uni-polar regions of negative polarity flux. As the magnetic field maps could have been measured up to $\pm30$ minutes around the detection of the parent IRIS burst, we note that these magnetic field maps may not be entirely representative of the exact magnetic field structure at the time of potential reconnection as, for example, cancellation could have occurred removing evidence of any bi-poles (see for example, \citealt{Nelson16}) for the three uni-polar events. Additionally, the spatial resolution of the SDO/HMI instrument may be insufficient for resolving all of the magnetic field elements at these locations (as discussed by \citealt{Reid16}) which could cause us to detect a uni-polar region when a bi-polar region was actually present. It is clear, however, that the majority of parent IRIS bursts with areas between $1$ arcsec$^2$ and $4$ arcsec$^2$ occur co-spatial to regions of opposite polarity magnetic field in the lower solar atmosphere, supporting the assertion that these events could be driven by magnetic reconnection.

\section{Conclusions}
\label{Conclusions}

In this article, we used a slightly modified (see Sect.~\ref{Obs_mod}) version of the algorithm originally applied in \citet{Kleint22} in order to investigate whether changes in the measurable properties of IRIS burst profiles occurred as their host ARs evolved. We studied $42$ dense IRIS rasters which each sampled one of seven host ARs, with each AR being sampled at least four times. We found:
\begin{itemize}
\item{The maximum total unsigned magnetic flux within an AR, and the complexity of the structuring of that flux, may be important as an indicator of whether IRIS burst profiles are present. The two ARs with the lowest peak total unsigned magnetic fluxes contained far fewer IRIS burst profiles than the five ARs with the highest unsigned magnetic fluxes. Specifically, AR $11871$ (with a peak unsigned magnetic flux of $5.38\times10^{20}$) was found to contain no IRIS burst profiles in the four rasters which sampled it, whilst AR $11856$ (with a peak unsigned magnetic flux of $8.60\times10^{20}$ was found to contain only $25$ IRIS burst profiles ($0.62$ \%\ of the total number of IRIS burst profiles studied here) across the five rasters which sampled it. The age of the AR was not found to influence the number of IRIS burst profiles returned by the algorithm, with the two youngest ARs ($11916$ and $11871$) being found to contain the most and least IRIS burst profiles, respectively.}
\item{In total, $4019$ IRIS burst profiles were returned by the algorithm from these $42$ rasters, with these being found to belong to $752$ parent IRIS bursts. These parent IRIS bursts had a mean area of $0.31$ arcsec$^2$, with $93.09$ \%\ of the identified events having areas below $1$ arcsec$^2$ (centre panel of Fig.~\ref{Fig_stats}). We note that the mean area should not be over-interpreted due to the large standard deviation from this dataset ($0.72$ arcsec$^2$) but it can be used to show that these events are typically small-scale. The frequencies and areas of these features did not monotonically increase or decrease as the host ARs evolved but instead varied in a seemingly randomly manner from raster to raster. The largest parent IRIS burst found through this analysis had an apparent area of $8.98$ arcsec$^2$ (corresponding to $151$ pixels). The overall spatial frequency of IRIS burst profiles across the studied ARs was calculated to be $0.00069$ arcsec$^{-2}$, well below the coverage of spicules, for example.}
\item{IRIS burst profiles are typically confined to localised regions (around $100$\arcsec$\times60$\arcsec\ for the two ARs studied here) close to the cores of ARs (as is displayed in Fig.~\ref{Fig_loc}). The average spatial separation between the centre-of-mass of all IRIS burst profiles within a raster and the individual IRIS burst profiles themselves appeared to increase with time in ARs $11850$ and $11916$ (the only ARs which contained more than $100$ IRIS burst profiles in multiple rasters), suggesting IRIS bursts occurred across a larger area as their host ARs evolved. This increase in area potentially corresponds to the growth of the ARs themselves. These results suggest that one must be very careful when selecting appropriate targets for co-observations of IRIS burst-Ellerman bomb pairs with ground-based telescopes with limited FOVs, e.g. with the Daniel K. Inouye Solar Telescope (DKIST; \citealt{Rimmele20, Rast21}).}
\item{The spectral properties (e.g. peak intensities, spectral locations of the peak intensities) of IRIS burst profiles varied in a non-monotonic way in datasets sampling ARs $11850$ and $11916$ which contained more than $100$ IRIS burst profiles, and then only slightly, through time. On top of this, the automatically detected IRIS burst profiles were found to be spread across a large number of the reference spectra used by the detection algorithm with, on average, unique reference spectra equalling $47.57$ \%\ of the total number of IRIS burst profiles in any given raster. Additionally, no individual reference spectrum was found to be the best fit for more than $7.34$ \%\ of the IRIS burst profiles returned for any given raster. These combined result indicate that no `typical' IRIS burst spectra exists for the rasters studied here.}
\item{$93$ \%\ of all parent IRIS bursts with areas between $1$ arcsec$^2$ and $4$ arcsec$^2$ occurred co-spatial to bi-poles in the solar photosphere. This result further supports the assertion that these events are driven by magnetic reconnection in the lower solar atmosphere (\citealt{Peter14, Vissers15, Hansteen17}).}
\end{itemize} 
Expanding this research to include a larger number of ARs (i.e. those sampled after 2014) will allow us to make further inferences about IRIS bursts in the future.

\begin{acknowledgements}
CJN is thankful to ESA for support as an ESA Research Fellow. LK gratefully acknowledges funding via a SNSF PRIMA grant. IRIS is a NASA small explorer mission developed and operated by LMSAL with mission operations executed at NASA Ames Research Center and major contributions to downlink communications funded by ESA and the Norwegian Space Centre. SDO/HMI data provided courtesy of NASA/SDO and the HMI science team. This research has made use of NASA’s Astrophysics Data System Bibliographic Services.
\end{acknowledgements}

\bibliographystyle{aa}
\bibliography{UV_ARs}

\newpage
\begin{appendix}
\section{Summary of observations and results}

\begin{table}[!hbt]
\centering
\caption{Summary of the $42$ datasets studied in this article.} 
\begin{tabular}{| c | c | c | c | c | c | c | c | c | c | c |}
\hline
\bf{AR} & \bf{Date} & \bf{Start (UT)} & \bf{End (UT)} & \bf{x$_c$\arcsec} & \bf{y$_c$\arcsec} & \bf{Exp (s)} & \bf{Rasters} & \bf{BPs} & \bf{Bs} & \bf{OBSID} \\ \hline
AR 11850 & 2013-09-24 & 05:09:45 & 05:29:12 & -327 & 92 & 2 & 1 & 109 & 39 & 4000254145 \\ \hline
AR 11850 & 2013-09-24 & 11:44:43 & 12:04:10 & -265 & 88 & 2 & 1 & 181 & 21 & 4000254145 \\ \hline
AR 11850 & 2013-09-24 & 15:39:43 & 15:59:10 & -264 & 70 & 2 & 1 & 5 & 3 & 4000254145 \\ \hline
AR 11850 & 2013-09-25 & 06:39:43 & 06:59:10 & -88 & 85 & 2 & 1 & 41 & 11 & 4000254145 \\ \hline
AR 11850 & 2013-09-25 & 11:09:43 & 11:29:10 & -53 & 85 & 2 & 1 & 312 & 46 & 4000254145 \\ \hline
AR 11850 & 2013-09-26 & 05:59:43 & 06:58:04 & 130 & 90 & 2 & 3 & 258 & 28 & 4000254145 \\ \hline
AR 11850 & 2013-09-26 & 11:09:43 & 11:29:10 & 178 & 98 & 2 & 1 & 26 & 9 & 4000254145 \\ \hline
AR 11850 & 2013-09-27 & 05:24:32 & 05:48:21 & 327 & 82 & 2 & 1 & 156 & 45 & 3800254046 \\ \hline
AR 11850 & 2013-09-27 & 06:24:43 & 06:44:10 & 335 & 78 & 2 & 1 & 56 & 15 & 4000254145 \\ \hline \hline
AR 11856 & 2013-10-09 & 05:31:44 & 06:31:59 & 177 & 35 & 8 & 1 & 0 & 0 & 3820259446 \\ \hline
AR 11856 & 2013-10-09 & 09:35:44 & 10:35:44 & 258 & 36 & 8 & 1 & 13 & 5 & 3820259446 \\ \hline
AR 11856 & 2013-10-10 & 05:00:58 & 06:01:13 & 376 & 29 & 8 & 1 & 7 & 4 & 3820259446 \\ \hline
AR 11856 & 2013-10-11 & 05:20:57 & 06:22:17 & 622 & 74 & 8 & 1 & 5 & 2 & 3820259646 \\ \hline
AR 11856 & 2013-10-11 & 10:21:27 & 11:22:47 & 627 & 51 & 8 & 1 & 0 & 0 & 3820259646 \\ \hline \hline
AR 11871 & 2013-10-18 & 13:42:56 & 17:14:43 & -244 & 257 & 30 & 1 & 0 & 0 & 3820013446 \\ \hline
AR 11871 & 2013-10-18 & 19:50:30 & 23:22:17 & -194 & 263 & 30 & 1 & 0 & 0 & 3820013446 \\ \hline
AR 11871 & 2013-10-19 & 17:55:46 & 18:56:16 & -48 & 225 & 15 & 1 & 0 & 0 & 3820009446 \\ \hline
AR 11871 & 2013-10-20 & 12:15:00 & 17:48:53 & 91 & 273 & 15 & 3 & 0 & 0 & 3820011446 \\ \hline \hline
AR 11909 & 2013-12-03 & 06:24:38 & 06:57:55 & -38 & -317 & 4 & 1 & 30 & 17 & 3800256046 \\ \hline
AR 11909 & 2013-12-03 & 07:54:14 & 08:14:07 & -76 & -273 & 2 & 1 & 126 & 20 & 3800254046 \\ \hline
AR 11909 & 2013-12-03 & 11:14:38 & 11:47:55 & 18 & -305 & 4 & 1 & 43 & 12 & 3800256046 \\ \hline
AR 11909 & 2013-12-03 & 17:54:44 & 18:14:37 & 27 & -272 & 2 & 1 & 64 & 18 & 3800254046 \\ \hline
AR 11909 & 2013-12-03 & 21:14:38 & 21:47:55 & 109 & -314 & 4 & 1 & 54 & 8 & 3800256046 \\ \hline
AR 11909 & 2013-12-04 & 03:25:09 & 03:45:02 & 162 & -306 & 2 & 1 & 57 & 15 & 3800254046 \\ \hline
AR 11909 & 2013-12-05 & 18:44:51 & 19:18:11 & 495 & -327 & 4 & 1 & 54 & 11 & 3800256196 \\ \hline
AR 11909 & 2013-12-05 & 20:22:31 & 20:55:51 & 453 & -297 & 4 & 1 & 6 & 2 & 3800256196 \\ \hline \hline
AR 11916 & 2013-12-06 & 07:34:51 & 08:08:11 & 36 & -223 & 4 & 1 & 715 & 65 & 3800256196 \\ \hline
AR 11916 & 2013-12-06 & 15:49:31 & 16:22:51 & 97 & -222 & 4 & 1 & 120 & 25 & 3800256196 \\ \hline
AR 11916 & 2013-12-06 & 22:20:41 & 22:54:01 & 158 & -226 & 4 & 1 & 288 & 58 & 3800256196 \\ \hline
AR 11916 & 2013-12-07 & 03:09:49 & 03:43:09 & 202 & -225 & 4 & 1 & 236 & 39 & 3800256196 \\ \hline
AR 11916 & 2013-12-07 & 11:14:51 & 11:48:11 & 273 & -220 & 4 & 1 & 283 & 45 & 3800256196 \\ \hline
AR 11916 & 2013-12-07 & 22:46:41 & 23:20:01 & 390 & -216 & 4 & 1 & 273 & 59 & 3800256196 \\ \hline
AR 11916 & 2013-12-09 & 07:10:51 & 07:44:11 & 633 & -216 & 4 & 1 & 73 & 2 & 3800256196 \\ \hline
AR 11916 & 2013-12-09 & 21:56:54 & 22:30:14 & 721 & -214 & 4 & 1 & 64 & 21 & 3800256196 \\ \hline \hline
AR 12104 & 2014-07-04 & 11:40:30 & 15:10:55 & -114 & -237 & 30 & 1 & 42 & 14 & 3824263396 \\ \hline
AR 12104 & 2014-07-05 & 23:00:30 & 02:30:55$^*$ & 203 & -238 & 30 & 1 & 72 & 32 & 3824263396 \\ \hline
AR 12104 & 2014-07-07 & 23:35:30 & 03:05:55$^*$ & 562 & -226 & 30 & 1 & 54 & 7 & 3824263396 \\ \hline
AR 12104 & 2014-07-08 & 19:26:13 & 22:56:38 & 702 & -220 & 30 & 1 & 0$^\dag$ & 0 & 3824263396 \\ \hline \hline
AR 12139 & 2014-08-15 & 07:08:03 & 08:10:51 & -439 & 120 & 8 & 1 & 94 & 34 & 3800258196 \\ \hline
AR 12139 & 2014-08-15 & 22:36:09 & 02:09:34$^*$ & -305 & 112 & 30 & 1 & 10 & 7 & 3880012196 \\ \hline
AR 12139 & 2014-08-16 & 21:17:49 & 22:20:37 & -106 & 103 & 8 & 1 & 45 & 12 & 3800258196 \\ \hline
AR 12139 & 2014-08-17 & 21:44:03 & 22:46:51 & 105 & 104 & 8 & 1 & 47 & 1 & 3800258196 \\ \hline \hline
\end{tabular}

\tablefoot{General information about the $42$ datasets studied here, including: The AR sampled; the date the observation started on; the start time of the observation; the end time of the observation ($^*$ indicates the following day); the initial x-coordinate of the centre of the FOV; the initial y-coordinate of the centre of the FOV; the exposure time per raster step; the number of rasters sampled during the observation; the number of IRIS burst profiles (BPs) detected using automated methods ($^\dag$ indicates that IRIS burst profiles were subsequently apparent in our manual analysis); the number of parent IRIS bursts (Bs) identified; and the OBSID of the experiment.}
\label{Tab_overview}
\end{table}

\label{appendix}
\end{appendix}

\end{document}